\documentclass[12pt]{article}
 \usepackage{amsfonts}
\usepackage{amsmath}
\usepackage{youngtab}
\usepackage{booktabs}
\usepackage{mathrsfs}
\usepackage{here}

\newcommand{\be}{\begin{equation*}}
\newcommand{\ee}{\end{equation*}}
\newcommand{\bea}{\begin{eqnarray*}}
\newcommand{\eea}{\end{eqnarray*}}

\newcommand{\e}{\epsilon}
\newcommand{\ie}{{\em i.e.}~}
\newcommand{\eg}{{\em e.g.}~}

\def\scr{\mathscr}

\def\punkt{\,\,.}
\def\komma{\,\,,}

\def\a{\alpha}

\def\e{\varepsilon}
\def\g{\gamma}

\def\m{\mu}
\def\n{\nu}
\def\l{\lambda}

\def\G{\Gamma}

\def\Z{\mathbb{Z}}

\def\R{\mathbb{R}}

\def\x{$\times$}

\newcommand\fr[1]{\hbox{$\genfrac{}{}{}{}{1}{#1}$}}

\def\dual{{\star}}
\def\M{{\scr M}}

\def\tP{\tilde\Pi}
\def\EE{{\scr E}}
\def\tEE{\tilde{\scr E}}
\def\tE{\tilde E}
\def\tF{\tilde F}

\def\w{\wedge}
\def\*{\partial}

\begin{document}

\begin{center}\Large{\textbf{M-branes on U-folds}}\end{center}

\begin{center}{Martin Cederwall}\end{center}

\catcode`\@=11
\begin{center}{\footnotesize{Fundamental Physics}}\\
{\footnotesize{Chalmers Univ. of Technology}}\\
{\footnotesize{SE-41296 G\"oteborg, Sweden}}\\
{\footnotesize{martin.cederwall@chalmers.se}}\end{center}
\catcode`\@=\active

\vskip4mm

\noindent{\footnotesize{\underbar{\textit{Abstract:}} We give a preliminary 
discussion of how
the addition of extra coordinates in M-theory, which together with the
original ones parametrise a U-fold, can serve as a tool for
formulating brane dynamics with manifest U-duality. The redundant
degrees of freedom are removed by generalised self-duality constraints
or calibration conditions made possible by the algebraic structure of
U-duality. This is the written version of an invited talk at the 7th 
International Workshop ``Supersymmetries and Quantum Symmetries'',
Dubna, July 30--August 4, 2007.}}

\vskip6mm

\noindent $D=11$ supergravity, dimensionally reduced on a torus $T^n$, 
has a global symmetry 
$E_{n(n)}$. The scalar fields parametrise the coset 
$E_{n(n)}(\Z)\backslash E_{n(n)}/K(E_{n(n)})$
\cite{Dualisation1,Dualisation2,HullTownsend,UDualityReview}.
The series has a natural continuation to the infinite-dimensional
cases of $E_9$, $E_{10}$ and $E_{11}$, although it is clear that the
content of the latter two is larger than the fields in the
dimensionally reduced theory. They have been proposed to actually
describe M-theory, either in a picture where space or space-time is
emergent \cite{DHN,DHNReview,LivingReview},
or hypothetically already in $D=11$ \cite{West:2001as,Englert:2003zs}.

With only gravity, the internal vielbein parametrises 
$SL(n,\Z)\backslash GL(n)/SO(n)$, the size and shape of the torus.
The symmetry enhancement comes from ``mixing'' of gravitational
and tensorial fields (dualised or not).
On reduction to $d=3$, even pure gravity gives a symmetry 
enhancement---the graviphotons are dualised to scalars 
and become part of a $SL(n+1)$
``vielbein'' (Ehlers symmetry) \cite{Ehlers}.

\begin{table}[H]
\begin{tabular}{ccc}
\toprule
  $n$&$E_{n(n)}$&$K(E_{n(n)})$\\
\midrule
$2$&$SL(2)\times\mathbb{R}$&$SO(2)$\\
$3$&$SL(3)\times SL(2)$&$SO(3)\times SO(2)$\\
$4$&$SL(5)$&$SO(5)$\\
$5$&$Spin(5,5)$&$(Spin(5)\times Spin(5))/\mathbb{Z}_2$\\
$6$&$E_{6(6)}$&$USp(8)/\mathbb{Z}_2$\\
$7$&$E_{7(7)}$&$SU(8)/\mathbb{Z}_2$\\
$8$&$E_{8(8)}$&$Spin(16)/\mathbb{Z}_2$\\
\bottomrule
\\
\noalign{\textit{Table 1. U-duality groups and their maximal compact
  subgroups.}}
\end{tabular}
\end{table}

Let us sketch how U-duality arises in a simple example, namely $n=4$.
We divide the 11-dimensional coordinates $X^M$ in $x^\m$, $\m=1,\ldots,7$,
coordinates on the uncompactified 7-dimensional space-time, and
$y^m$, $m=1,\ldots,4$, coordinates on the torus $T^4$.

The massless bosonic fields are
\begin{center}
\begin{table}[H]
\begin{tabular}{ll}
$g_{\m\n}$:&metric, singlet;\\
$C_{\m\n\l}\leftrightarrow \tilde C_{\m\n}$, $C_{\m\n p}$:&2-forms in {\bf 5}
of $SL(5)$;\\
$g_{\m n}$, $C_{\m np}$:&1-forms in {\bf 10} of $SL(5)$;\\
$g_{mn}$, $C_{mnp}$:&scalars in $SL(5)/SO(5)$.\\
\end{tabular}
\end{table}
\end{center}
\noindent This matches with the decomposition of representations when
$SL(5)\rightarrow SL(4)\times \R$:
\vskip6pt
\Yboxdim6pt
\begin{center}
\begin{table}[H]
\begin{tabular}{cccl}
\yng(1)&${\bf 5}$&$\rightarrow$&${\bf 4}_{1/5}\oplus{\bf 1}_{-4/5}$\\
\yng(1,1)& 
\raise3pt\hbox{${\bf 10}$}&\raise3pt\hbox{$\rightarrow$}&
\raise3pt\hbox{${\bf 6}_{2/5}\oplus{\bf 4}_{-3/5}$}\\
\yng(2,1,1,1) 
&\raise9pt\hbox{${\bf 24}$}&\raise9pt\hbox{$\rightarrow$}&
\raise9pt\hbox{${\bf 4}_{1}\oplus({\bf 15}\oplus{\bf 1})_0
\oplus\bar{\bf 4}_{-1}$}\\
\end{tabular}
\end{table}
\end{center}
\noindent As we consider dimensional reduction,
the fields do not depend on the coordinates of $T^n$.

The global symmetry $E_{n(n)}$ (\eg\ $E_{5(5)}\approx SL(5)$) contains
rigid transformations changing the shape and size of the torus as well as
the C-field on the torus. This can be seen as diffeomorphisms and 2-form 
gauge transformations with parameters linear in coordinates.
This suggests the enlargement of the set of coordinates to a ``coordinate
representation'' of $E_{n(n)}$ \cite{Hull:2007zu,West:2007mh}. 
Note that this is the same
representation as the set of massless 1-forms. The coordinate
representation is given by the natural enlargement of the
``graviphotons''.

The following table gives the coordinate representations case by case, 
together with their reduction to
representations of $SL(n)$:

\begin{table}[H]
\begin{tabular}{cccccccc}
\toprule
  $n$&$E_{n(n)}$&repr.&$P^m$&$Z_{mn}$&$Z_{m_1\ldots m_5}$
               &$Z_{m_1\ldots m_6}$&\\
\midrule
$2$&$SL(2)\times\mathbb{R}$&$2\oplus1$&\x&\x&&&\\
$3$&$SL(3)\times SL(2)$&$(3,2)$&\x&\x&&&\\
$4$&$SL(5)$&$10$&\x&\x&&&\\
$5$&$Spin(5,5)$&$16$&\x&\x&\x&&\\
$6$&$E_{6(6)}$&$27(\oplus1)$&\x&\x&\x&(\x)&\\
$7$&$E_{7(7)}$&$56$&\x&\x&\x&\x&\\
$8$&$E_{8(8)}$&$248$&\x&\x&\x&\x&+more...\\
\bottomrule
\\
\noalign{\textit{Table 2. Coordinate representations.}}
\end{tabular}
\end{table}

Hull \cite{Hull:2007zu} realised that the enlarged internal spaces
could be used to describe, and geometrise, 
classes of non-geometric solutions to
M-theory. In a situation where the remaining space-time is
topologically non-trivial, the complete space can be taken as a bundle
of the extended internal space over space-time where the allowed
holonomies are in the discrete U-duality group.
 
U-duality acts linearly on the coordinates of torus with extended coordinates, 
but not on $T^n$. (Consider \eg\ a T-duality on $T^1$ with 
$R\leftrightarrow R^{-1}$. If one instead interchanges two circles with
radii $R$ and $R^{-1}$ the patching is geometrical.)

The aim here is to initiate a search for 
formulations of brane dynamics using the extended
coordinates.
This gives a manifestly U-duality symmetric formulation of brane dynamics,
and can provide a geometric description of branes in situations that 
are not geometric in the unextended formalism.
There must be constraints on the branes, so that the dependence on the
extra coordinates is eliminated in a proper way.
Note also that branes of different dimensionalities transform into each other
under U-duality. This means that they are described by the same brane 
on the extended space, but some of its directions may be hidden in the
extra coordinate directions. 

Let us for a little while focus on and review the analogous questions
for T-duality \cite{Hull:2006va}. 
T-duality transformations form a subgroup $SO(n-1,n-1)$ of the U-duality
group. It is a perturbative symmetry of string theory on $T^n$.
(Probably inspired by Hitchin's generalised complex geometry,) Hull
proposed that $T^n$ should be enlarged to $T^{2n}$, on which T-duality
acts linearly, in order to geometrise non-geometric solutions of 
string theory, where transition functions contain non-geometric 
T-duality transformations.
He also gave a formulation of string dynamics on the extended space.

The metric and $B$-field parametrise an element of 
$SO(n,n)/(SO(n)\times SO(n))$.
Let us call the vector index of $SO(n,n)$ (the tangent index) $M$ and 
the flat vector indices of $SO(n)\times SO(n)$ $a$ and $a'$.
$$
\EE_M{}^a=\left[\begin{array}{c}
E^{ma}\\
F_m{}^a\\
\end{array}\right]
\komma\qquad
\tEE_M{}^{a'}=\left[\begin{array}{c}
\tE^{ma'}\\
\tF_m{}^{a'}\\
\end{array}\right]\komma
$$

$$
\begin{array}{cccccc}
\hfill E^{ma}&\!\!=\!\!&\fr{\sqrt2}e^{ma}\komma\hfill
\hfill \tE^{ma'}&\!\!=\!\!&-\fr{\sqrt2}e^{ma'}\komma\hfill \\
\hfill F_m{}^a&\!\!=\!\!&\fr{\sqrt2}(e_m{}^a-B_{mn}e^{na})\komma\quad\hfill
\hfill \tF_m^{a'}&\!\!=\!\!&\fr{\sqrt2}(e_m{}^{a'}+B_{mn}e^{na'})\punkt\hfill\\
\end{array}
$$
The invariant metrics are 
\bea
L_{MN}&=&(\EE\EE^t-\tEE\tEE^t)_{MN}=\left[
  \begin{array}{cc}0&1\\
    1&0\\
  \end{array}\right]\komma\cr
&&\cr
&&\cr
G_{MN}&=&(\EE\EE^t+\tEE\tEE^t)_{MN}=
\left[\begin{array}{ccc}g^{mn}&&g^{mp}B_{pn}\\
-B_{mp}g^{pn}&&g_{mn}-B_{mp}g^{pq}B_{qn}\\
\end{array}\right]\punkt\cr
\eea

How is string dynamics realised on the extended space, the T-fold?
We will give a somewhat simplified account, forgetting quantum issues.
Let us for simplicity forget about the uncompactified directions (they are of course important, but the technical issue is to get rid of unwanted dependence of the ``too many'' compactified directions). It is straightforward to insert
\eg\ the graviphoton field later.

We call the coordinates $Z^M=(Y_m,X^m)$.
The pullbacks of the frame 1-forms to the string world-sheet are
\bea
\Pi^a&=&dX^me_m{}^a+(dY_n-dX^mB_{mn})e^{na}\komma\cr
\tP^{a'}&=&dX^me_m{}^{a'}-(dY_n-dX^mB_{mn})e^{na'}\punkt\cr
\eea
In order to reduce the number of degrees of freedom to half one imposes
duality constraints
\bea
\dual\Pi^a&=&\Pi^a\komma\cr
\dual\tP^{a'}&=&-\Pi^{a'}\cr
\eea
(which may also be written as $\dual dZ^M=L^{MN}G_{NP}dZ^P$).
So $\Pi$ and $\tP$ are left- and right-moving, respectively.
The (anti-)selfduality implies that the equations of motion
are automatically satisfied.

Quantum calculations need to use holomorphic factorisation. The partition
function has been shown to agree with string theory \cite{Berman:2007vi}.

We would now like to do something similar for branes of M-theory. 
There are several issues.
Branes with $p>1$ are nonlinear. There is no conformal gauge. 
If there is a duality relation involved, dualisation 
with which world-volume metric? There is essentially one 
candidate: the pullback $\G_{ij}$ of the metric 
$G_{MN}$ on the extended space determined by
the $E_{n(n)}/K(E_{n(n)})$ vielbein.

Dualisation of some 1-form $\Pi^A$ on a $(p+1)$-dimensional brane gives a 
$p$-form. One needs some invariant tensor $c$ to be able to write
$$
\dual_\G\Pi^A=c^A{}_{A_1\ldots A_p}\Pi^{A_1}\wedge\ldots\wedge\Pi^{A_p}\punkt
$$
 Which are the possible invariant tensors? Let us make a list.
We assume that the tensor $c_{AA_1\ldots A_p}$ is totally antisymmetric, and
look for singlets under $K(E_{n(n)})$ in the $p+1$-fold product 
$\wedge^{p+1}R$ of the coordinate representation. (The table is not complete.)

\begin{table}[H]
\begin{tabular}{ccccccccccccc}
\toprule
  $n$&$R_{E_{n(n)}}$&$R_{K(E_{n(n)})}$ repr.
  &$p=$&$2$&$3$&$4$&$5$&$6$&$7$&$8$&$9$&$10$\\
\midrule
$4$&$10$&$10$&&\x&&&&\x&&&&\\
$5$&$16$&$(4,4)$&&&\x&&&&\x&&&\\
$6$&$27$&$27$&&&&\x&&&&\x&&\\
$7$&$56$&$28\oplus\bar{28}$&&&&&\x&&&&\x&\\
$8$&$248$&$120\oplus128$&&&&&&\x&...&?&&\\
\bottomrule
\\
\noalign{\textit{Table 3. Series of $K(E_{n(n)})$-invariant
    antisymmetric tensors.}}
\end{tabular}
\end{table} 

%(The table is not complete...)

%{Detailed example: M2 instanton, $n=4$}
In order to show that the idea works, we will do one example in
detail, the case of an M2 instanton with $n=4$.
The term ``instanton'' just means that we consider a 
euclidean membrane spanning {\it only}
internal directions. This simplifies things: we don't need to worry
how the movement of the membrane in the uncompactified directions enter
into the world-volume metric. (This has to be investigated, of course.)

%\underbar{Notation:}

The U-duality group is $SL(5)$. 
Coordinates are $Z^\M=Z^{MN}=(X^m,Y^{mn})$ where 
$M=1,\ldots,5$, $m=1,\ldots,4$.
The metric on the repr. $10$ is $G_{MN,PQ}=\fr2 G_{M[P}G_{Q]N}$, where
$G_{MN}$ is the metric on the repr $5$.
The $SL(5)$ vielbein is parametrised as
$$
E_M{}^A=\left[\begin{array}{cc}
e^{1/3}&-e^{1/3}C^ne_n{}^a\\
0&e^{-1/3}e_m{}^a\\
\end{array}\right]\komma
$$
where $C^m=\fr6\e^{mnpq}C_{npq}$.
This gives the metric
$$
G_{MN}=\left[\begin{array}{cc}
g^{1/3}(1+C^pC^qg_{pq})&-C^pg_{pn}\\
-g_{mp}C^p&g^{-1/3}g_{mn}\\
\end{array}\right]\punkt
$$
The induced metric on the world-volume is 
$\G_{ij}=\fr2\*_iZ^{MN}\*_jZ^{PQ}G_{MP}G_{NQ}$.
Splitting $Z^{MN}$ into $Z^{0m}=X^m$ and $Z^{mn}=g^{1/3}Y^{mn}$, one gets
\bea
\G_{ij}&=&\fr2\*_iZ^{mn}\*_jdZ^{pq}G_{mp}G_{nq}
+2\*_{(i}Z^{0n}\*_{j)}Z^{pq}G_{0p}G_{nq}\cr
&&+\*_iZ^{0m}\*_jZ^{0n}(G_{00}G_{mn}-G_{0m}G_{0n})\cr
&=&\fr2\*_iY^{mn}\*_jY^{pq}g_{mp}g_{nq}
-2\*_{(i}X^m\*_{j)}Y^{np}C_ng_{mp}\cr
&&+\*_iX^m\*_jX^n((1+C^2)g_{mn}-C_mC_n)\punkt\cr
\eea
We want to use this metric in the dualisation, and see if we can find 
reasonable solutions of $Y$ in terms of $X$ to the duality relation
$$
\dual_\G dZ^{MN}=\a dZ^{MP}\w dZ^{NQ}G_{PQ}\punkt
$$

An Ansatz for the solution may be
$$
\dual_\g dY^{mn}=adX^m\w dX^n+b\dual_\g dX^{[m}C^{n]}\komma
$$
where dualisation is with $\g_{ij}$, the pullback to $T^4$ of $g_{mn}$.
Inserting this Ansatz into the metric gives
$$
\G_{ij}=(1+a^2+(1+\fr2b)^2C^2)\g_{ij}-(1+\fr2b)^2C_iC_j\punkt
$$
We choose $b=-2$, otherwise the ordinary membrane eqs. of motion can not
be recovered. Then $\G_{ij}=(1+a^2)\g_{ij}$, and acting on a 1-form,
$\dual_\G=\sqrt{1+a^2}\dual_\g$
Now we insert this into the duality relations, which then read
\bea
\a^{-1}\sqrt{1+a^2}\dual_\g dY^{mn}&=&dY^{mp}\w dY^n{}_q
       +2dX^{[m}\w dY^{n]p}C_p+(1+C^2)dX^m\w dX^n\komma\cr
-\a^{-1}\sqrt{1+a^2}\dual_\g dX^m&=&dY^{mp}\w dX_p+dX^m\w dX^pC_p\punkt\cr
\eea
The two equations must be consistent with each other and with the
Ansatz, which gives two equations,
$$
\a=\genfrac{}{}{}{}{\sqrt{1+a^2}}{2a}
=\genfrac{}{}{}{}{a}{\sqrt{1+a^2}}\komma
$$ 
with the solutions $a=\pm1$, $\a=\pm\fr{\sqrt2}$.
This is quite nontrivial, and depends on several cancellations.
It shows that the duality relation provides a U-duality
covariant description of the membrane on $T^4$. 

Note that the duality relation implies that the equations of motion
following from 
$$
S=\int d^3\xi\sqrt\G
$$
are satisfied. There is, and should not be, a separate WZ term,
since the $C$-field is contained in the metric on the enlarged space.

For the future, one should consider branes moving also in the uncompactified
directions, with couplings to all background fields, and also
different values of $n$. Supersymmetrisation is presumably straightforward.

%{Branes with tensor fields --- some preliminaries}

We would like to make some comments about branes with vector or 
tensor fields.
How do world-volume vector or tensor fields, such as the vector potential
on D-branes, or the self-dual 2-form on the M5-brane, arise? Normally
such fields are identified with the value of a background tensor field
on the brane, but these fields are now unified with the metric. 

If $n$ is large enough, a U-duality rotation relates \eg\ 
an M2-brane and an M5-brane. Either the tensor field is not present in the 
dynamics on the extended space, but arises as a parametrisation of
the orientation of the brane in the extra directions, or it is made to
disappear for the M2 interpretation.

No concrete case has been worked through, and
we will instead take D-branes as an example, where a qualitative
description has been given by Hull \cite{Hull:2006va}. 
All D-branes on the doubled torus
are $n$-dimensional and span a light-like $n$-plane with respect to
the metric $L$. 
Which brane is ``seen'' depends on the choice of
polarisation, \ie, of the choice of the embedding of $GL(n)$ 
in $SO(n,n)$.

Consider an orientation where $k$ of the directions lie in the $X$
directions (\ie, the rank of $\*_iX^m$ is $k$). Then there are
$n-k$ transverse directions in $X$ and $k$ in $Y$. The extra ($Y$) transverse
directions are related to the vector field.
A bit more precisely:
The possible duality relations one can write down for an $n$-dimensional
brane are
\bea
\Pi^a&=&\pm\fr{(n-1)!}\e^a{}_{a_2\ldots a_n}\dual_\G(\Pi^{a_2}\w\ldots\w\Pi^{a_n})\komma\cr
\tP^{a'}&=&\pm\fr{(n-1)!}\e^{a'}{}_{a'_2\ldots a'_n}\dual_\G(\tP^{a'_2}\w\ldots\w\tP^{a'_n})
\punkt\cr
\eea
Take the case where the D-brane fills the $X$-space. Then $\*_iX^m$ is
non-degenerate, and one may try
\bea
\*_iY_m&\sim&f_{ij}
\e^{jj_2\ldots j_n}\e_{mm_2\ldots m_n}\*_{j_2}X^{m_2}\ldots\*_{j_n}X^{m_n}
+\*_iX^nB_{nm}\cr
&\sim&f_{ij}(\*X)^{-1}_m{}^j+\*_iX^nB_{nm}\punkt\cr
\eea

Suppressing the matrices $\*X$ (\ie, using a static gauge where it is
the unit matrix), one has
\bea
\Pi&=&1+f\komma\cr
\tP&=&1-f\komma\cr
\eea
and this is a (quite nontrivial) solution to the proposed duality relations.
It is also the correct deformation preserving the lightlikeness
of the orientation of the brane.
The dynamics of the vector field (eqs. of motion and Bianchi identities)
should be examined closer. Here it arises in the solution of the
generalised self-duality constraint. 
It is still not excluded that a complete 
formulation demands a field that lives intrinsically on the brane.
The corresponding investigation should also be carried on to M-branes
in cases where $n\geq6$.

%{Outlook}

Let us summarise, by giving an outlook, 
ranging from technical points to wild speculation:

There are many cases to be worked through in detail, 
especially the ones with world-volume
tensor fields. Coupling to all background fields has to be included,
but should be straightforward.
Supersymmetry and $\kappa$-symmetry should certainly be manageable, but
will take some work. It may be interesting to examine the special
cycles on the internal manifold from the viewpoint of supersymmetric
calibrations. 

Polarisations---choices of the physical subspace---correspond to 
embeddings of $SL(n)$ into $E_{n(n)}$. Pure spinors parametrise the 
corresponding cosets in the doubled formalism in the context of
T-duality (and play an important r\^ole in generalised complex geometry).
Is there a natural generalisations of pure spinors 
to the M-theoretic setting? (The answer 
is probably yes.)

A question that has bearing on the $E_{11}$ proposal, concerns the
``reality'' of the extra coordinates.
Is there a Borisov--Ogievetsky-like construction 
\cite{Borisov:1974bn} containing diffeomorphisms
of the torus and gauge transformations of tensor fields, without generating
``everything''? If this is possible, what is the relation 
to higher spin theory? Like higher spin theory, this would be a model
with massless higher spin fields, obtained by extending the
coordinates with some tensorial objects.

It should be noted that descriptions of
brane dynamics respecting global symmetries of
string theory or M-theory have been worked out earlier with other
methods (see \eg\ refs. \cite{Cederwall:1997ts,Cederwall:1997ab,Cederwall:1998xr,Cederwall:1999zf,Bengtsson:2004nj}). We expect such descriptions to be
reproduced by the present program when branes spanning only the
uncompactified direction are considered.

\vskip6pt
\noindent\underbar{\textit{Acknowledgements:}} The speaker is grateful 
for discussions with Bengt EW Nilsson and Peter West.

%\end{thebibliography}

\end{document}